\documentclass[11pt,a4paper,a4wide]{article}
\usepackage{amsfonts,amssymb,amsmath,amsopn,cite}
\usepackage{bm}
\usepackage{isolatin1}
\usepackage{graphicx}
\usepackage[german]{varioref}
\usepackage[dvips]{epsfig}
\usepackage{graphics}
\usepackage[dvips]{color}
\usepackage{color}
\usepackage{colordvi}
\usepackage{dsfont}
\usepackage{psfrag}
\begin{document}
\newcommand{\captionfonts}{\small}
\newcommand{\be}{\begin{eqnarray}}
\newcommand{\ee}{\end{eqnarray}}
\def\p#1#2{|#1\rangle \langle #2|}
\def\ket#1{|#1\rangle}
\def\bra#1{\langle #1|}
\def\refeq#1{(\ref{#1})}
\def\tb#1{{\overline{{\underline{ #1}}}}}
\def\im{\mbox{Im}}
\def\re{\mbox{Re}}
\def\nn{\nonumber}
\def\t{\mbox{tr}}
\def\sgn{\mbox{sgn}}
\def\Li{\mbox{Li}}
\def\P{\mbox{P}}
\def\d{\mbox d}
\def\i{\int_{-\infty}^{\infty}}
\def\ip{\int_{0}^{\infty}}
\def\mi{\int_{-\infty}^{0}}
\def\A{\mathfrak A}
\def\AA{{\overline{{\mathfrak{A}}}}}
\def\a{\mathfrak a}
\def\aa{{\overline{{\mathfrak{a}}}}}
\def\B{\mathfrak B}
\def\BB{{\overline{{\mathfrak{B}}}}}
\def\b{\mathfrak b}
\def\bb{{\overline{{\mathfrak{b}}}}}
\def\R{\mathcal R}
\def\dm{\mathfrak d}
\def\dd{{\overline{{\mathfrak{d}}}}}
\def\D{\mathfrak D}
\def\DD{{\overline{{\mathfrak{D}}}}}
\def\c{\mathfrak c}
\def\cc{{\overline{{\mathfrak{c}}}}}
\def\C{\mathfrak C}
\def\CC{{\overline{{\mathfrak{C}}}}}
\def\Or{\mathcal O}
\def\F{\mathcal F_k}
\def\N{\mathcal N}
\def\I{\mathcal I}
\def\S{\mathcal S}
\def\G{\Gamma}
\def\L{\Lambda}
\def\la{\lambda}
\def\g{\gamma}
\def\al{\alpha}
\def\s{\sigma}
\def\e{\epsilon}
\def\k{\kappa}
\def\ve{\varepsilon}
\def\te{\text{e}}
\def\rmi{\text{i}}
\def\max{\text{max}}
\def\str{\text{str}}
\def\tr{\text{tr}}
\def\tC{\text C}
\def\Fo{\mathcal{F}_{1,k}}
\def\Ft{\mathcal{F}_{2,k}}
\def\vs{\varsigma}
\def\l{\left}
\def\r{\right}
\def\up{\uparrow}
\def\down{\downarrow}
\def\u{\underline}
\def\ov{\overline}
\title{Crossover from  Anderson- to Kondo-like behavior:\\
Universality induced by spin-charge separation}
\author{Michael Bortz, Andreas Kl\"umper and Christian Scheeren}
\maketitle

\subsection*{Abstract}
The thermodynamics of a lattice regularized asymmetric Anderson impurity in a correlated host is obtained by an exact solution. The crossover from the Anderson- to the Kondo-regime is studied, thus making contact with predictions by scaling theory. On the basis of the exact solution, the transition to universal Kondo behavior is shown to be realized by a graduate separation of the energy scales of spin and charge excitations.  

\section{Introduction}
\label{intsec}
Homogeneous correlated one-dimensional electron systems have been studied intensively in recent
years and exact results have been obtained by as different techniques as
bosonization \cite{aff91,aff95,gog98}, conformal field theory (CFT) \cite{fran97}, and the
algebraic Bethe Ansatz (ABA) \cite{kor93}. Whereas bosonization and CFT deal
with the ground state and low-temperature properties of the continuum models,
the Quantum Transfer Matrix (QTM) technique has been developed \cite{kl93} for
the calculation of the thermodynamics of lattice models. The QTM combines the
Bethe Ansatz solution with the quantum inverse scattering method and a
Trotter-Suzuki mapping of the partition function on a classical
two-dimensional lattice. In this approach, the free energy of the model is encoded in a
set of finitely many non-linear integral equations (NLIE). All these methods
have been applied for example to the general spin-1/2 Heisenberg chain
\cite{kl93}, the $tJ$-model \cite{jue97tj} and the Hubbard model \cite{jue98,
  kl96}. At low temperatures, i.e. in the conformal limit, the different
methods yield identical results.  

The equivalence of continuum and lattice methods is much less investigated in the case of inhomogeneous systems, where a localized
impurity interacts with the host by $s$-wave scattering, so that the model is reduced to one dimension. First, the Kondo and the Anderson
models have been solved exactly by the coordinate Bethe Ansatz \cite{tsv83}, in
this case starting directly from the continuum model, i.e., a point-like impurity
coupled to a free host. This approach has been extended to
calculate thermodynamic properties by applying the Thermodynamic Bethe Ansatz
(TBA), which yields the free energy in terms of infinitely many coupled
NLIE. In the case of the spin-1/2 single channel Kondo model, the presence of
a stable infrared fixed point is the starting point for CFT-calculations
\cite{aff91, aff95}. These confirm results of the Bethe-Ansatz solution and go
beyond. 

Later, several schemes of including an impurity in the correlated host on a lattice accessible by the QISM have been found \cite{andr84,frahm,bar94}. However, the connection between the continuum and the lattice models in the conformal limit with a generic Kondo-like interaction has not been established until very recently \cite{bo04}. In \cite{bo04}, a lattice regularized asymmetric Anderson impurity in a correlated host has been obtained from QISM, and the free energy has been calculated exactly by the QTM-method. As a special case, the Kondo limit has been performed explicitly, yielding a conformally invariant bulk of free fermions interacting with a magnetic impurity through spin exchange. The results in the Kondo limit are identical with and extend those from the BA-solution, but they are derived from a finite number of NLIE. 

Here, we study the properties of the model proposed in \cite{bo04} in the interesting regime of crossover from Anderson- to Kondo-like properties. The model consists of a three-state impurity permitting single and zero occupation in a correlated fermionic host. The impurity interacts with its nearest host neighbor lattice sites by charge fluctuation and spin exchange. By employing the Poor Man's scaling approach proposed by Anderson \cite{and70} in the framework of the Kondo model and applied subsequently to the generic asymmetric Anderson model \cite{jef77,hal78}, three different regimes at low temperatures are qualitatively distinguished, namely the mixed valence regime (MV), the strong coupling regime (SC) and the local moment regime (LM). This qualitative description is confirmed quantitatively on the basis of the exact solution. Especially, the universal strong coupling limit is attained by energetically decoupling charge and spin excitations.  

% Especially, impurity models studied previously by QISM and TBA-techniques \cite{frahm, bar94} are contained in our formulation. Additionally, the host in our model covers the two-parameter-model \cite{jue97} and the $tJ$-model \cite{jue97tj}. We show that the question whether a crossover to a localized impurity takes place is decided by the impurity properties, independently of the bulk. 

The key ingredient of our approach is the one-parametric family of four-dimensional irreps of gl(2$|$1), from which a gl(2$|$1)-symmetric transfer matrix is constructed, yielding the impurity model in the Hamiltonian limit. We use different representations labeled by different interaction parameters in the host and impurity spaces. 
%From this point of view, our work follows up \cite{pfan96, grun99}, where the eigenvalues and eigenstates of the corresponding transfer matrix have been calculated. 
The motivations to consider gl(2$|$1) are twofold: On one hand, the limit
of free fermions is contained as a special case and is given by algebraic
functions (rather than by trigonometric functions like in the $XXZ$ model). On the other hand, su(2) is one of the even subalgebras of gl(2$|$1), and one expects that gl(2$|$1) symmetry can be reduced to su(2) symmetry in a certain limit of the parameters. This is indeed true for special parameter choices and constitutes the Kondo limit. 

This article is organized as follows. In the next section the model is defined, including a qualitative analysis in the Poor Man's scaling approach. The third section contains the quantitative study of the different regimes on the basis of the exact solution. The results are summarized in a phase diagram. The fourth section focuses on the Luttinger-liquid behavior of the host. In the last section, conclusions are drawn.

%%%%%%%%%%%%%%%%%%%%%%%%%%%%%%%%%%%%%%%%%%%%%%%%%%%%%%%%%%%%%
%%%%%Section2
%%%%%%%%%%%%%%%%%%%%%%%%%%%%%%%%%%%%%%%%%%%%%%%%%%%%%%%%%%%%%
\section{Definition of the model and qualitative analysis}
The Hamiltonian is defined as the logarithmic derivative of an inhomogeneous transfer matrix given in appendix \ref{appa}. It is the sum of the host and impurity contributions \cite{bo04},
\be
H&=& H_h+\P H_i \P+H_{ex}\label{mod}\\
H_h&=& -\sum_{k,\sigma} \l(2 D(\cos k-1)+\mu\r)c^\dagger_{k\sigma} c_{k\sigma} +H_{int}\label{hmod}\\
H_i&=& E_0-(E_0+\mu) n_d  +\sum_{k,\sigma}\l[V_k c^\dagger_{k\sigma}d_{\sigma}+h.c.\r]+\sum_{k,p,\sigma} J_{k,p}\l( c^\dagger_{p,\sigma} c_{k\ov\sigma} d^\dagger_{\ov\sigma}d_{\sigma} +c^\dagger_{p,\sigma} c_{k\sigma}n_{d\sigma}\r)\label{imod}\\
H_{ex}&=& \frac{h}{2}\sum_\sigma\l[\sigma d^\dagger_\sigma d_\sigma + \sum_{k} \sigma c^\dagger_{k\sigma}c_{k\sigma}\r]\nn,
\ee
where $\ov \sigma=-\sigma$. Interactions in the host are allowed which are
parameterized by a representation theoretic parameter $\al$ such that
$H_{int}= \Or(1/\al)$ \cite{bar95}; the limit $\al\to \infty$ corresponds to
free fermions. The limit $\al\to0$ yields the $tJ$-model, which can be seen by a canonical transformation \cite{bar95}. In \refeq{mod}, the impurity contribution is projected onto non-doubly occupied states by the projection operator $\P$. The energy of the non-occupied impurity state in \refeq{imod} is $E_0=-4 D \al^2/(4 u_0^2+\al^2)$. In the limit $\al\gg 1$, the coupling constants are given by
\be
V_k&=& \frac{\al^{3/2} \sqrt{D}}{2\l(u_0^2+\al^2/4\r)}\l(\rmi u_0\l(\te^{-\rmi k}-1\r) +\l(\te^{-\rmi k}+1\r)\r)\label{vk}\\
J_{k,p}&=& \frac{\al}{2(u_0^2+\al^2/4)} \l(\rmi u_0\l(\te^{\rmi k}-\te^{-\rmi p}\r) +\l(\te^{\rmi k}+\te^{-\rmi p}\r)\r)\label{jkp}.
\ee
Additionally to \cite{bo04}, where we focussed on the most important terms
even in $u_0$, the leading terms uneven in $u_0$ are included here.

The model \refeq{mod} describes an asymmetric Anderson model in a correlated
host, where the doubly-occupied impurity state is dynamically decoupled and
energetically suppressed. Interactions are tuned by $\al, u_0$, the bandwidth
$D$ and the chemical potential $\mu$. The homogeneous system, which consists
of $H_h$ alone, was studied in \cite{jue97, sak01}. To obtain a qualitative
understanding of the model, note that for the impurity model with $u_0=0$, the
thermodynamic quantities depend only on $T/D$ for fixed $\mu/D$. The meaning
of different choices of $D/\mu$ for the finite-temperature behavior is
revealed by studying the scaling properties following the Poor Man's approach,
\cite{and70,jef77,hal78}.  It consists of investigating the leading
renormalization of the coupling constants by the reduction of $D$ to
$D-|\delta D|=:D^{(eff)}$, integrating out states with energies
$|\epsilon|>D^{(eff)}$. This renormalization procedure is done in the limit
$T\ll D$, at sufficiently low temperatures compared to the bandwidth. The resonance width $\Delta(\omega)$ is defined by
\be
\Delta_{\omega}=\sum_{k=-\pi}^\pi\frac{\l|V_{k}\r|^2}{D} \delta(\omega-\e_k) n_k\nn\;,
\ee
where $\e_k=2 D(\cos k-1)+\mu$ is the bare one-particle dispersion relation of
the host, eq. \refeq{hmod}, and $n_k$ is the occupation of the energy level
$\e_k$. The resonance width has a maximum at the Fermi level, $\Delta_0\equiv
\Delta$, and decays to zero at the band edges. Let us consider the scaling properties of $\Delta$, and the energy difference between single and non-occupation, 
\be
\varepsilon_d:=-\mu-E_0\label{vedef}.
\ee 
By performing a canonical transformation, the hybridization terms in \refeq{imod} are eliminated to leading order, see \cite{bo04} for details. Let us first concentrate on the relations between $\delta D$ and $\delta \ve_d, \,\delta \Delta$ (constant prefactors are omitted):
\be
\delta \ve_d&\sim& -\Delta \delta \ln D\nn\\
\delta \Delta&\sim& \frac{\Delta\cdot \Delta_{2D}}{D}\delta \ln D\nn
\ee
From the last equation, $\Delta$ remains unrenormalized in the limit
$D\to\infty$. The first equation can then be integrated to give
$\ve_d^{(eff)}+\Delta\ln D^{(eff)}=\ve_d^*$, where $\ve_d^*$ is a scaling
invariant and $\ve_d^{(eff)}:=\ve_d(D^{(eff)})$. This means that
$\ve_d^{(eff)}$ grows logarithmically if $D^{(eff)}$ is reduced. Depending on
the relation of $\ve_d^{(eff)}$ to $D^{(eff)}$, one distinguishes between three cases: 
\begin{itemize}
\item[i)] $\ve_d^{(eff)}\in\l[-D^{(eff)},D^{(eff)}\r]$: Mixed valence (MV)
\item[ii)] $\l|\ve_d^{(eff)}\r|\gg D^{(eff)}$: Strong coupling (SC)
\item[iii)] Completely filled band such that $D^{(eff)}=0$ or $D^{(eff)}$ is
  not defined: Local moment (LM). 
\end{itemize}
The first two cases are realized by $\mu$ sufficiently large to ensure a
finite filling of the band but $\mu<\mu_c$, where the critical potential
is calculated below from the exact solution as $\mu_c=4 D(\al+1)/\al$. In the
MV case, both $\mu$ and $D$ are finite, so that $\ve_d^{(eff)}$ is situated in
the band or near the band edges with the impurity orbital retaining its
triplet degeneracy. The signature in the susceptibility is a crossover from
$T\chi(T)=1/6$ at temperatures $T>\Delta$ to the Fermi liquid property
$\chi(T= 0)=$const., with a non-universal constant value. It originates in a
non-integral impurity occupation number at low temperatures; that is why this
case is referred to as the ''mixed valence'' (MV) regime. 

The conditions of the second case are met for $\mu<\mu_c$ in the limit
$\mu\to\infty$ at fixed $\mu/D$. Single occupation of the impurity is enforced
and $\ve_d^{(eff)}$ moves outside the band upon scaling, so that only virtual
charge fluctuations between the impurity and the host occur. This crossover
takes place at an effective bandwidth $D^*$ given by
$D^*:=-\ve_d^{(eff)}=D^{(eff)}$, i.e. $D^*-\Delta \ln D^*=-\ve_d^*$. The exact
solution presented below shows that $D^*$ decreases with decreasing
$D/\mu$. Since here, we are dealing with temperatures $T\ll D$, where $D$ is the initial bandwidth, one can neglect the momentum dependence of $J_{k,p}$ and take the momenta at the Fermi points $J_{k,p}=J_{k_F, p_F}$. The linear order in $u_0$ contributes only if $p_F=k_F$, so that we define $J_{k_F, k_F}=:J$. Now the Poor Man's approach is applied to the spin exchange terms \refeq{jkp}, which again can be done by a canonical transformation eliminating these terms to the leading order. This results in 
\be
\delta J\sim - J^2\delta \ln D\label{scalj},
\ee
from which a new scaling invariant 
\be
T_K:= D\te^{-1/J}=D^{(eff)} \te^{-1/ J^{(eff)}}\label{tk1}
\ee
is deduced, where $J$ is given by \refeq{jkp}. In the limit $|u_0|\gg 1$, the sign of the spin exchange depends on both the sign of $k_F$ and $u_0$. For $u_0<0$, spin-exchange is antiferromagnetic at the right Fermi point $k_F>0$ and vice versa. From \refeq{scalj} it follows that antiferromagnetic exchange dominates over the ferromagnetic term in the scaling procedure, so that from \refeq{jkp}, the antiferromagnetic exchange constant is given by $J\sim |\sin k_F|/|u_0|$ for $|u_0|\gg 1$. Comparison with \refeq{tk1} leads to $\l|u_0\r|\sim\ln D$. The susceptibility is expected to show a crossover from $T\chi(T)=1/6$ to $1/6<T\chi(T)<1/4$, ending at $\chi(0)= 1/(2\pi T_K)$, which is the universal value in the strong coupling (SC) regime \cite{wil75}.

A further reduction of the initial bandwidth $D$, or an increase of the
chemical potential at fixed $D$ such that $\mu>\mu_c$ moves $D^*\to 0$ and
drives the system into the third case. Then at low temperatures, the chemical potential lies outside the band. No interactions between the impurity and the host are possible, so $T\chi(T)$ changes from 1/6 to 1/4 from high to low temperatures. This is the local moment (LM) regime. 

Summarizing the results from scaling arguments, one finds that $u_0$ plays a
crucial role in the crossover to the strong-coupling limit and is determined
by the renormalization of the antiferromagnetic spin-coupling constant
$J$. The additional $u_0$-dependence of $\Delta$, see \refeq{vk}, is neglected in the Poor Man's scaling; however, it has to be included in order to render the model integrable.

Let us briefly compare our model with the generic asymmetric Anderson model \cite{and61},
\be
H_i=\ve_d\sum_\sigma n_{d,\sigma} +U n_{d\up}n_{d\down}+\sum_{k,\sigma} V_{k} c^\dagger_{k\sigma} d_{\sigma}+h.c.\label{andmod}
\ee
where $U\gg |\ve_d|,\Delta$ such that double occupation of the impurity site
is excluded. Contrary to \refeq{andmod}, the model \refeq{mod} does not
include backscattering \cite{pun97}; it contains a bare spin-exchange term,
which is generated in \refeq{andmod} by a canonical transformation
\cite{sch66} in the limit $U/\Delta\gg 1$. These modifications of \refeq{andmod} are necessary in order to
make our model integrable {\em on a lattice}, but do not alter the physics:
The model \refeq{andmod} also exhibits the three regimes discussed above in
the appropriate asymmetric limit $U\gg \ve_d, \Delta$ \cite{hal78}, with the further possibility to tune $\ve_d$ in \refeq{andmod} such that the impurity is completely depleted at low temperatures. With respect to the three regimes discussed in this work, \refeq{mod} and \refeq{andmod} are equivalent. 
%%%%%%%%%%%%%%%%%%%%%%%%%%%%%%%%%%%
%%Section3
%%%%%%%%%%%%%%%%%%%%%%%%%%%%%%%%%%
\section{Exact solution}
The rather qualitative picture of the preceding section is made more precise by the exact solution yielding the free energy of the host per lattice site $f_h$ and the impurity $f_i$ in terms of suitably chosen auxiliary functions $\B=1+\b$, $\BB=1+\bb$, $\C=1+\c$, cf. appendix \ref{appa}.
\be
\ln \b(v)\!\!&=&\!\!-\l[\Phi * \ln \frac{\C}{\c}\r](v+\rmi /2) +\beta h/2 +[k*\ln \B](v)-[k*\ln
  \BB](v+\rmi)\label{bv3}\\
\ln \bb(v)\!\!&=&\!\!\l[ \Phi*\ln \frac{\C}{\c}\r](v-\rmi /2) -\beta h/2 +[k*\ln \BB](v)-[k*\ln
  \BB](v-\rmi)\label{bbv3}\\
\ln \c(v)\!\!&=&\!\! -\beta
  D(\al+1)\Psi_{\al}(v)+\beta\mu-\l[k*\ln \frac{\C}{\c}\r](v)\nn\\
& &+[\Phi*\ln
  \B](v-\rmi/2)-[\Phi*\ln \BB](v+\rmi/2)\label{cv3},
\ee
where the integration kernels and driving term 
\be
\Phi(v)&:=&\frac{\rmi}{2\sinh \pi v}\,,\;\Psi_\al(v):=\frac{\al}{v^2+\al^2/4}\label{psidef}\\
k(v)&=&\frac{1}{2\pi} \i \frac{\te^{-|k|/2}}{2\cosh k/2} \te^{\rmi k v}\d k\nn
\ee
have been defined. The convolutions are defined as $[f*g](v):= \i f(v-w)g(w)\d w$. With the solution of \refeq{bv3}-\refeq{cv3} one obtains
\be
f_i&=&- \mu-T\l[\ln \frac{\C}{\c}*k+ Dk\r](u_0) -\frac{T}{2\pi} \i \frac{[\ln \B\BB](v)}{\cosh (u_0-v)}\d v\label{fi}\\
f_h&=& -2\mu+2 D -T\l[\Psi_{\al}*\ln \frac{\C}{\c}\r](0)\,\d v\label{fh}\nn.
\ee 
In this section we discuss the three regimes distinguished above. The NLIE are therefore solved analytically near complete filling at zero temperature and numerically for arbitrary parameters. 

At low temperatures the driving term of $\ln \c$, eq. \refeq{cv3}, gets
arbitrarily large and determines the location of zeroes of $\ln \c$. It is
convenient to define the quantities $\bar \mu$, $\mu_c$ by 
\be
\bar \mu&:=&\mu_c-\mu\nn\\
\mu_c&:=& 4 D(\al+1)/\al\label{mucdef}.
\ee
It is shown below that: 
\begin{itemize}
\item[i)] No zeroes of $\ln \c$ occur for $\mu>\mu_c$. This case corresponds to complete filling of the band, where the impurity is in the LM regime.
\item[ii)] $\ln \c$ acquires two zeroes $\pm\ln\L_\c$ for $\mu<\mu_c$. For
  finite 
\be
|u_0|=(\ln D)/\pi\label{param},
\ee
$|u_0|\lesssim \L_\c$, the MV regime is realized. From \refeq{param}, justified rigorously in section
  \ref{scsec} and the
  $D$-dependence of $\L_\c$ calculated below \refeq{lamc}, one identifies
  $\ve_d^{(eff)}\sim |u_0|$ and $D^{(eff)}\sim \L_\c$.  
\item[iii)] If in the MV regime $u_0$ in \refeq{param} is chosen such that
  $|u_0|\sim \L_\c$ (corresponding to $D^{(eff)}\equiv D^*$ defined in the
  previous section), the crossover to the SC regime takes place. This regime
  is completely realized in the
  limit $\mu\to\infty$ at fixed $\mu/D$.
\end{itemize}
%%%%%%%%%%%%%%%%%%%%%%%%%%%%%%%%%
%%Subsection LM regime
%%%%%%%%%%%%%%%%%%%%%%%%%%%%%%%%%
\subsection{LM regime}
\label{lmsec}
For $\bar \mu <0$ at low temperatures, $\ln \C\equiv\ln \c$. Then the
eqs. \refeq{bv3}, \refeq{bbv3} are trivially solved by $\ln \b=\beta h=-\ln
\bb$, such that the free energy \refeq{fi} is essentially that of an uncoupled
local moment. Thus $\chi_i(T\to 0)=1/(4 T)$ diverges at lowest
temperatures. At high temperatures, $\ln\b$, $\ln \bb$, $\ln \c$ approach
their asymptotic values \cite{bo04}, such that the susceptibility reaches the value $\chi(T\to \infty)=1/(6 T)$. The crossover between these two extremes is shown in fig. \refeq{susmu300} with the values $D=1, 30$ at $\mu =300$. 
%
%
%%%%%%%%%%%%%%%%%%%%%%%%%%%%%%%%%
%%Subsection LM regime
%%%%%%%%%%%%%%%%%%%%%%%%%%%%%%%%%
\subsection{MV regime}  
\label{mvsec}
To gain analytical insight at $T=0$, scaled auxiliary functions and their zeroes are defined:
\be
\ve_\b=T\ln \b&;&  \ve_\c=T\ln \c\nn\\
\ve_\b\l(\pm \L_\b\r)=0 &;&\ve_\c\l(\pm \L_\c\r)=0\nn
\ee
Since in the low-temperature limit $\ln \bb\to -\infty$ at finite $h$, the function $\ln \BB$ vanishes. 
Following \cite{ess96}, one can apply Wiener-Hopf-techniques to the linearized NLIE (appendix \ref{appb}) at low temperatures for $|\bar \mu,h|\ll 1$. Performing analogous calculations as in \cite{ess96}, one can show that:
\be
\L_\c^2&=& \frac{\al^3}{16D(\al+1)} \l(\bar\mu -\frac{a\gamma}{\pi}  +\frac{\al^{3/2}\ln 2}{3\pi \l(D(\al+1)\r)^{1/2}}\bar\mu^{3/2}\r)+\Or\l(a \bar \mu^{1/2}, \bar \mu^{5/2}\r)\label{lamc}\\
\ve_\c(v)&=&-D(\al+1)\Psi_\al(v)+\mu-\frac{a\gamma}{\pi}\cosh(\pi v)-\frac{\al^{3/2}}{6\pi \l(D(\al+1)\r)^{1/2}}k(v)\bar\mu^{3/2}\nn\\
& &+\frac{a \gamma \al^{3/2}}{8\pi^2\l(D(\al+1)\r)^{1/2}}\bar \mu+\Or\l(a \bar \mu^{3/2}, \bar \mu^{5/2}\r)\label{vec} \\
\gamma&:=&\l.\int_{-\L_\c}^{\L_\c} \te^{\pi w} \ve_\c(w)\d w\r|_{h=0}=  \frac{\al^{3/2}}{3D \l(D(\al+1)\r)^{1/2}} \bar\mu^{3/2}+\Or\l(\bar \mu^{5/2}\r)\label{gamdef}\\
&=& \frac{64}{3} \frac{\al+1}{\al^3}\L_\c^3+\Or\l(\L_\c^5\r)\label{gam2} .
\ee
Here $a:=\frac{h^2}{2\gamma^2}$ and we defined a quantity $\gamma$ whose meaning will become clear later. The above expressions are valid for small $\bar \mu(>0)$ and $h$, i.e. near complete filling and small magnetization. Note that $\L_\c^2$ is not defined for $\bar \mu <0$, and neither is $\g$. 

Inserting the above results into the expression for the free energy \refeq{fi}, \refeq{fh} one obtains:
\be
f_i&=&-\mu-\frac{\g \cosh(\pi u_0)}{\pi} a -\frac{\al^{3/2} k(u_0)}{2\pi \l(D(\al+1)\r)^{1/2}} \bar\mu^{3/2}+\frac{\al^{3/2} \g k(u_0)}{8\pi^2 \l(D(\al+1)\r)^{1/2}}a\bar \mu\nn\\
n_i(h=0)&=& 1- \frac{3 \al^{3/2} k(u_0)}{4\pi \l(D(\al+1)\r)^{1/2}} \bar\mu^{1/2}=1-\delta n_i \label{ni}\label{chii}\\
\chi(h=0)&=& \frac{\cosh(\pi u_0)}{\pi\g} -\frac{\delta n_i}{6\pi \gamma}\nn\\
f_h&=& -2\mu+\frac{2 D(\al+1)}{\al} -\frac{2}{3\pi} \l(\frac{\al}{D(\al+1)}\r)^{1/2} \bar \mu^{3/2}+\frac{\g}{\pi^2} \l(\frac{\al}{D(\al+1)}\r)^{1/2} a\bar\mu^{1/2}\nn\\
n_h(h=0)&=& 2-\frac{1}{\pi} \l(\frac{\al}{D(\al+1)}\r)^{1/2} \bar \mu^{1/2}=:2-\delta n_h \label{n_h}\\
\chi_h(h=0)&=& \frac{\delta n_h}{\pi \gamma}\label{chih}\; .
\ee
At constant $\bar \mu$, the impurity occupation and susceptibility are tuned by the parameter $u_0$: For finite $u_0$, the impurity site is not fully occupied, and neither is the host since $\bar \mu >0$. The susceptibilities are expressed by the deviation from complete occupation, which justifies the term ``mixed valence''. For $u_0\to\infty$, single occupation on the impurity is enforced, whereas the host is away from complete filling. Especially, for $u_0=(\ln D)/\pi$, the impurity susceptibility is independent of $D$. It is shown below that this choice of $u_0$ leads to the Kondo regime for all temperatures. 

The high temperature asymptotics are $\chi_i(T\to\infty)=1/(6 T)$ and $\chi_h(T\to\infty)=1/(8 T)$, corresponding to three and four available states, respectively.

\begin{figure}
\begin{center}
\vspace{-0cm}
 \includegraphics[angle=-90,scale=0.4]{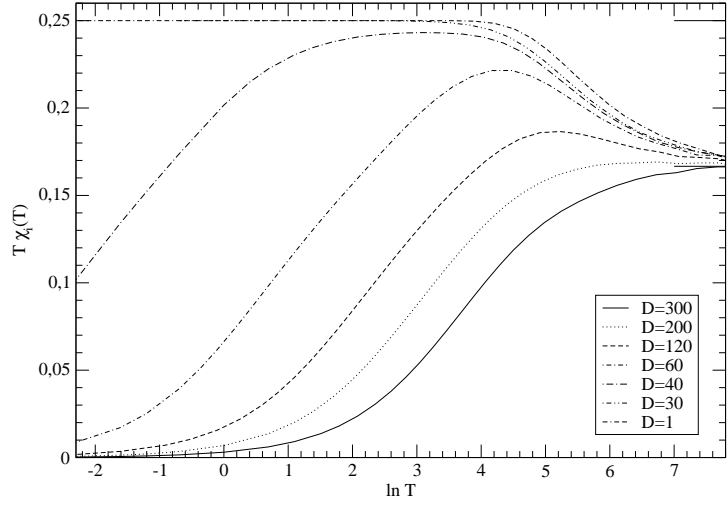}
\caption{Effective Curie constant of the impurity at $\mu=300$ for different $D$. The curves with $D=1,30$ exhibit LM-behavior, the one with $D=300$ corresponds to the MV-regime. In the other cases shown, signatures of the SC-regime are discernible. At high temperatures, the asymptotic values $1/6$ and $1/4$ are given.}
\label{susmu300}
\end{center}
\end{figure}

The crossover of $\chi_i(T)$ from low to high temperatures at arbitrary values of $\bar \mu$ is shown in figs. \ref{susmu300} and \ref{susdom2o5}. 
%
%
%%%%%%%%%%%%%%%%%%%%%%%%%%%%%%%%%
%%Subsection SC regime
%%%%%%%%%%%%%%%%%%%%%%%%%%%%%%%%%
\subsection{SC regime}  
\label{scsec}
By increasing $\mu,D$ at fixed $\mu/D$ and with $u_0=(\ln D)/\pi$, the crossover from the MV to the SC regime is performed. In terms of the zeroes $\pm \L_\c$ of $\ln \c$, this is the case when $\L_\c\ll (\ln D)/\pi$. Then by substituting $v=-x+(\ln D)/\pi$, the convolution in the driving terms in \refeq{bv3}, \refeq{bbv3} is approximated by
\be
[\ln\frac{\C}{\c}*\Phi](v)=-\beta \te^{-\pi x}T_K\label{app},
\ee
where 
\be
\beta T_K D:= \i \te^{\pi w} \ln [\C/\c](w) \d w=:\kappa \label{tkdef}.
\ee
The formula for $T_K$ in \cite{bo04} is an approximation of \refeq{tkdef}. From \refeq{tkdef}, it follows that $T_K$ is a monotonous function of $D/\mu$. By choosing $\pi u_0=\mp\ln D$ in \refeq{psidef}, one arrives at the equations yielding the free energy of a Kondo impurity,
\be
\ln \b(x)&=& -\te^{ x}  +\beta h/2+[k*\ln
\B](x)-[k*\ln\BB](x+\rmi\pi-\rmi \e)\label{bcon}\\
f_{i,spin}&=&-\frac{T}{2\pi} \i \frac{[\ln \B\BB](x)}{\cosh(x+\ln T/T_K)}\d x\label{konfen},
\ee
and $\BB=\l.\B^*\r|_{h\to-h}$. Thus, the functions $\B,\BB$ are decoupled from $\C$. This decoupling rests on \refeq{app}, which is valid for $\l[\ln \c'\r](\pm \L_F)\gg 1$ and $\L_F\ll (\ln D)/\pi=u_0$ at fixed $\mu/D$. The constant $T_K$ which appears in \refeq{konfen} is the scaling invariant expected from Poor Man's scaling in the Kondo regime. This becomes evident by writing 
\be
T_K=T \kappa/D=T\kappa \exp[- \pi |u_0|]\label{tku0},
\ee
where from \refeq{tkdef}, $\kappa \propto D/T$. The freedom in the choice of the sign of $u_0$ corresponds to the appearance of antiferromagnetic exchange with the right or left movers in \refeq{imod}, \refeq{jkp}, following the sign of $u_0$. By comparing \refeq{tkdef} with \refeq{gamdef}, one sees that $\g=D T_K$. 

\begin{figure}
\begin{center}
\vspace{-1.5cm}
\includegraphics[angle=-90,scale=0.4]{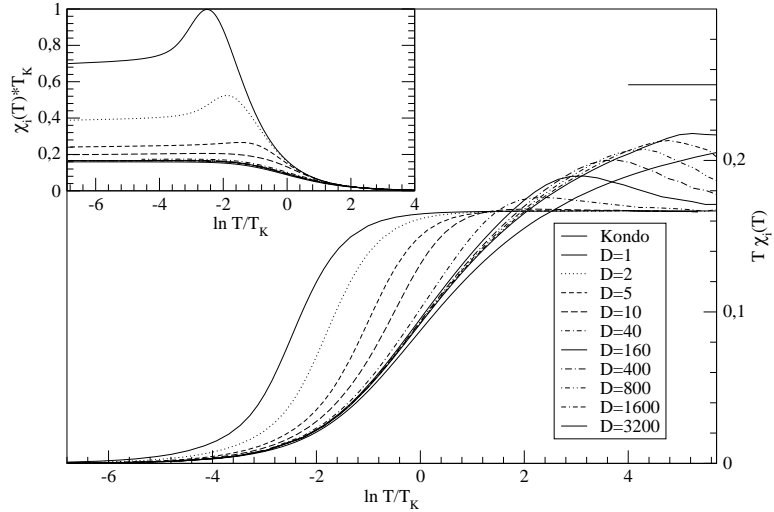}
\caption{Impurity susceptibility $\chi_i$ and $T\cdot \chi_i$ at constant
  $D/\mu=2/5$ illustrating the crossover from Anderson to Kondo behavior. At
  high temperatures, the asymptotic values $1/4$ and $1/6$ are indicated. For
  the ratio $D/\mu=2/5$, the zeroes of $\ln \c$ are found numerically to be
  $\pm\L_F=0.893$. So $u_0<\L_F$ corresponds to $D<16$. Note the different scaling behaviors for $u_0\gtrless \L_F$.}
\label{susdom2o5}
\end{center}
\end{figure}

If $D,\mu \gg 1$ but still finite, from the NLIE an approximate expression for $\ln \C$ is obtained, 
\be
\ln \C(v)=\ln \l[\l(1+\te^{\beta h/2-\beta  \phi_\c(v)/2+\beta \mu}+\te^{-\beta h/2-\beta  \phi_\c(v)/2+\beta \mu}\r)/\l(\B^{(\infty)}\BB^{(\infty)}\r)\r]\label{app1},
\ee
where $\phi_\c$ is defined in \refeq{phic} in appendix \ref{appa} and
$\B^{(\infty)}=1+\exp(\beta \mu +\beta h/2)$, $\BB^{(\infty)}=1+\exp(\beta \mu -\beta h/2)$. Then from \refeq{fimp} in appendix \ref{appa}, 
\be
f_{i,charge}=-\frac{T}{\pi}\i \frac{\ln \C(w)}{(w-u_0)^2+1}\d w\label{app2},
\ee
which is the free energy of an effectively free impurity with a certain
density of states, each state permitting no and single occupation. The
approach to the Kondo susceptibility is depicted in fig.~\ref{susdom2o5} for
constant $D/\mu=2/5$.

\begin{figure}
\begin{center}
\vspace{-1.0cm}
\includegraphics[angle=-90, scale=0.4]{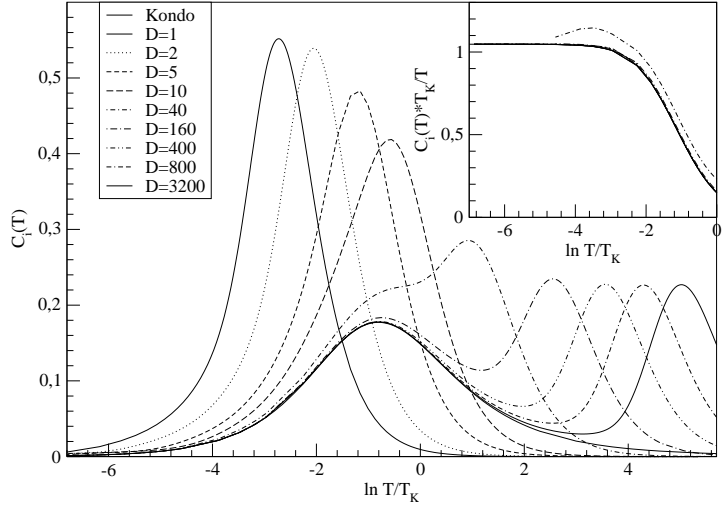}
\caption{Impurity specific heat at constant $D/\mu=2/5$ in the crossover region from Anderson to Kondo behavior. The inset shows the linear $T$ coefficient at low temperatures, the Kondo value being $\pi/3$. The peak corresponding to charge excitations shifts to higher temperatures. For the ratio $D/\mu=2/5$, the zeroes of $\ln \c$ are found numerically to be $\pm\L_F=0.893$. So $u_0<\L_F$ corresponds to $D<16$. Note the different scaling behaviors for $u_0\gtrless \L_F$. }
\label{sphdom2o5}
\end{center}
\end{figure}

Apart from the susceptibility, it is instructive to consider the specific heat
at constant $\mu$, shown in fig. \ref{sphdom2o5}. The approach of the Kondo
limit reveals two different scaling regimes. In a first step, for
$u_0<\Lambda_\c$, spin and charge excitations are not decoupled, but occur on
the same energy scale. This is shown by the single peak in
fig. \ref{sphdom2o5} for low enough $D$. With increasing $D$, the shift $u_0$
increases; once $u_0>\Lambda_\c$, the charge excitations are scaled to higher
and higher energies $\sim D$, whereas the new universal scale $T_K$ (only
dependent on the ratio $D/\mu$) appears as characteristic scale for the spin
excitations.  For $u_0\gg \L_\c$, spin- and charge excitations are
energetically well separated as described by \refeq{konfen}, \refeq{app2}. 

\begin{figure}
\begin{center}
\vspace{-0cm}
\includegraphics[angle=-90, scale=0.4]{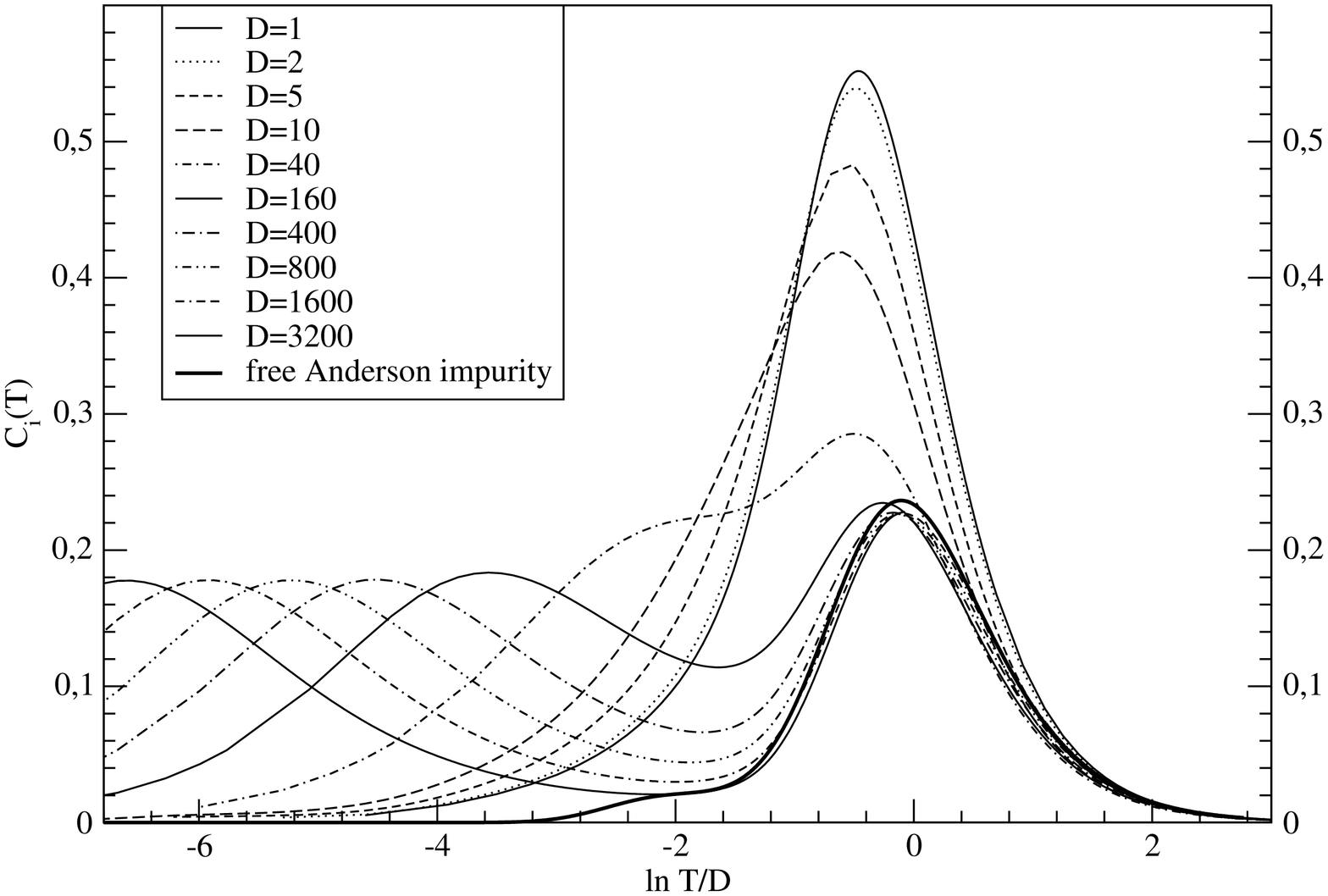}
\caption{Impurity specific heat at constant $D/\mu=2/5$ in the crossover region from Anderson to Kondo behavior, with the temperature scale $T/D$. Also shown is the specific heat derived from \refeq{app1}, \refeq{app2}, that is an effectively free three-state impurity, with parameters $D=3200$, $\mu=8000$.}
\label{sphdom2o5d}
\end{center}
\end{figure}
%
%%%%%%%%%%%%%%%%%%%%%%%%%%
%%%Phase diagram
%%%%%%%%%%%%%%%%%%%%%%%%%%
\subsection{Phase diagram}
\begin{figure}
\begin{center}
\vspace{-0cm}
 \includegraphics[scale=0.7]{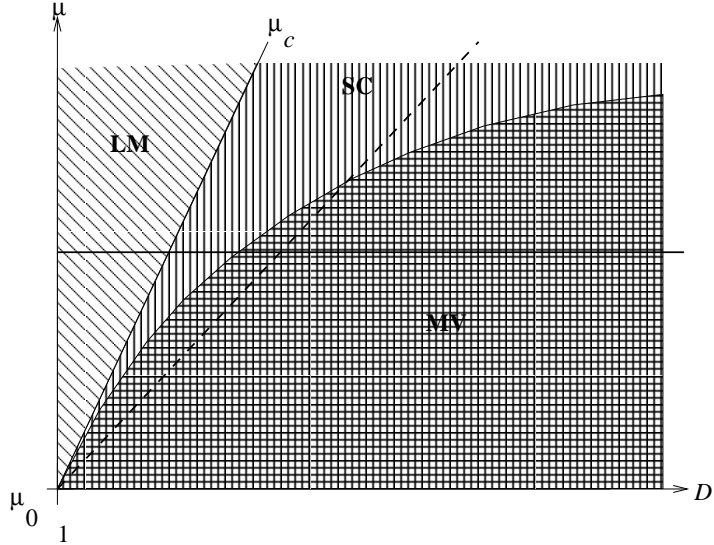}
\caption{Phase diagram showing the three different regimes of the
  impurity (diagonally hatched: LM, only vertically hatched: SC, vertically and horizontally
  hatched: MV) at low temperatures. The axes begin at $D=1$, $\mu_0:=\mu_c(D=1)=4(\al+1)/\al$. The line between MV and SC is the center of a transition region
  between these two regimes, determined by the condition $u_0=\ln D/\pi$. The impurity behaviour along the horizontal
  (diagonal dashed) line
  at $\mu=$const. ($\mu/D=$const.) is illustrated in fig. \ref{susmu300}
  (\ref{susdom2o5}-\ref{sphdom2o5d}). In the host, only two phases are
  present, LM (diagonally hatched) and MV (vertically hatched, that is MV and
  SC of the impurity). In the latter region, the
  host behaves like a Luttinger-liquid, whereas it is completely filled for $\mu>\mu_c$.}
\label{phasediag}
\end{center}
\end{figure}

The results of the previous sections \ref{lmsec}-\ref{scsec} are summarized in
figure \ref{phasediag}. For low temperatures, the three different impurity regimes are sketched in the
$\mu-D$-plane with $u_0=\ln D/\pi$. The LM-regime is distinguished from the
other two by the straight line $\mu=\mu_c$; for $\mu_c$, see \refeq{mucdef}. The SC- and MV-regimes are
separated by the condition $|u_0|=\ln \L_\c$. From \refeq{lamc}, 
\be
\mu=4 D(\al+1)/\al\l(1-\frac{4}{\al^2\pi^2}\ln^2D\r)\label{mud}
\ee
for $|\L_\c|\ll 1$ and $D\gtrsim 1$. Note that in \refeq{mud}, $\mu\to\mu_c$
for $D\to 1$. In the other extreme, at $|\L_\c|\gg 1$, the driving terms in
the linearized equations are expanded appropriately, yielding
\be
\L_\c^2&=&D(\al+1)\al/\mu\nn\\
\mu&=&D\pi^2(\al+1)\al/\ln^2D,\;\mbox{for }u_0=\L_\c\label{mud2}.
\ee 
Eq. \refeq{mud2} shows that the SC-regime can always be obtained for
$\mu<\mu_c$, $\mu/D$ fixed in the limit $D\to\infty$. Also shown in the phase
diagram \ref{phasediag} are the regions illustrated by figures
\ref{susmu300}-\ref{sphdom2o5d}. 

The SC regime can only be realized for the impurity. As far as the host is
concerned, the choices $\mu>mu_c$ ($\mu<\mu_c$) lead to a completely (not completely) filled
band. In the second case, the host behaves as a Luttinger liquid at low
temperatures, cf. appendix \ref{appb}. 
%

%%%%%%%%%%%%%%%%%%%%%%%%%%
%%%Subsection Host 
%%%%%%%%%%%%%%%%%%%%%%%%%
\subsection{Luttinger-liquid-behaviour of the host}
Since $u_0=0$ in the host, the only temperature scale is the bandwidth $D$, and all thermodynamic functions depend only on $T/D$ at constant $D/\mu$. In the LM-regime of the impurity, the band is completely filled, allowing for no excitations of the host. Away from complete filling, the host shows Luttinger-liquid-behavior for $\beta D\gg1$, as shown in appendix \ref{appb},
\be
f_h= -T^2\frac{\pi^2}{6}\l( \frac{1}{v_c}+\frac{1}{v_s}\r)\nn
\ee
 with distinct spin- and charge velocities $v_s$ and $v_c$. This corresponds
 to a linearized energy-momentum dispersion in \refeq{hmod}, the host interactions still being finite. In the Kondo limit, i.e. $D\to\infty$ at
 constant $D/\mu$, the Luttinger-liquid regime of the host is valid for all
 temperatures, including, as a special case, the free fermion limit at
 $\al\to\infty$. Thus the model presented here exhibits Kondo-like behavior
 even in the presence of an interacting bulk. In
\cite{fro95,fro96} arguments of CFT have been presented in order to decide whether generic singlechannel Kondo-like behavior is possible in an interacting host or not. The answer to this was found to depend on the host's properties. Further
studies of our model with techniques of conformal field theory are expected
to yield additional criteria to answer that question.

%%%%%%%%%%%%%%%%%%%%%%%%%%
%%%Conclusion
%%%%%%%%%%%%%%%%%%%%%%%%%
\section{Conclusion}
We have studied an asymmetric Anderson model in which a three-state impurity permits charge and
spin exchange with an interacting bulk on a lattice. For this model, we
distinguished a mixed valence, a strong coupling and a local moment regime
with the help of scaling arguments. This distinction was made quantitative by
the exact calculation of the free energy contribution of both the impurity and
of the host. On the basis of the exact solution, the three different regimes
have been studied both analytically and numerically. Analytically, a new
universal energy scale for excitations was deduced. Its gradual emergence was
illustrated by numerical calulations. The numerical solution illustrated the
emergence of a new energy scale, namely the Kondo temperature $T_K$, in the
strong coupling limit. In the Kondo limit, this scale is separated by an
increasingly large energy from the temperature on which charge fluctuations
are excited. Our model can be generalized to a four-state impurity, which also
permits double occupation. The derivation of the Hamiltonian and its analysis,
especially the comparison with the finite-$U$ Anderson model, are projects for future work. 

%%%%%%%%%%%%%%%%%%%%%%%%%%%%%%%%%%%%%
%%Appendix A
%%%%%%%%%%%%%%%%%%%%%%%%%%%%%%%%%%%%%
\appendix
\section{Appendix A}
\label{appa}
In this appendix we sketch the construction of the gl(2$|$1)-symmetric transfer matrix, from which the Hamiltonian \refeq{mod} is obtained, and the derivation of the NLIE \refeq{bv3}-\refeq{cv3}. 

Let $V^{(\al)}$ be the module of the 4-dimensional irrep of gl(2$|$1), labeled by the parameter $\al$. For $\al=0$, this is the three-dimensional fundamental irrep. Following \cite{pfan96,grun99}, we define matrices \\ $R^{(\al, \al')}\in$End$\l(V^{(\al)}\otimes V^{(\al')}\r)$, where the first (second) space is referred to as ``auxiliary'' (``quantum'') space. These matrices satisfy the graded Yang-Baxter-Equation
\be
 \l[R^{(\al,\al')}(u)\r]^{\beta, \g}_{\beta', \g'}\, \l[R^{(\al'',\al')}(v)\r]^{\delta, \g'}_{\delta', \g''}\,
\l[R^{(\al'',\al)}(v-u)\r]^{\delta', \beta'}_{\delta'',
  \beta''}\,(-1)^{(p[\delta]+p[\delta'])p[\beta']}\nn\\
\qquad =\l[R^{(\al'',\al)}(v-u)\r]^{\delta, \beta}_{\delta',\beta'}\, \l[R^{(\al'',\al')}(v)\r]^{\delta', \g}_{\delta'', \g'}\,
\l[R^{(\al,\al ')}(u)\r]^{\beta', \g'}_{\beta'',
  \g''}(-1)^{(p[\delta']+p[\delta''])p[\beta']}\label{ybedef}\; .
\ee   
Here a grading is assigned to the basis vectors through the parity function $p$, namely $p[1]=p[4]=0;\, p[2]=p[3]=1$. The matrix $R^{(\al,\al')}$ is written as a sum of projectors,
\be
R^{(\al,\al')}(v)=1+\frac{1}{q(v)} \check P_1^{(\al,\al')}-\frac{1}{p(v)} \check P_3^{(\al,\al')}\nn\;,
\ee
where $p(v)=1/2+v/(\al+\al'+2)$, $q(v)=-1/2+v/(\al+\al')$ and $\check P_1, \,\check P_3$ are projectors from
$V^{(\al)}\otimes V^{(\al')}$ onto gl(2$|$1) modules with highest weights
$(0,0|\al+\al')$ and $(-1,-1|\al+\al'+2)$ respectively. They are given explicitly in \cite{grun99,pfan96}. Note that $R^{(\al,\al')}(v)\,R^{(\al',\al)}(-v)=1$. Furthermore, consider the set of matrices
\be
\l[\ov
R^{(\al,\al')}(u)\r]^{\alpha,\beta}_{\gamma,\delta}=(-1)^{p[\delta](p[\g]+p[\alpha])}
\l[ R^{(\al,\al')}(-u)\r]^{\g,\beta}_{\al,\delta}\label{rbardef}\; ,
\ee 
which also satisfy a YBE. From $R$ and $\ov R$, monodromy matrices $T$ and $\ov T$ are constructed by multiplying $(L+1)$-many $R$-matrices in auxiliary space,
\be
T(u)&=&R_{a,L}^{(\al_h,\al_h)}(u)R_{a,L-1}^{(\al_h,\al_h)}(u)\ldots
R_{a,1}^{(\al_h,\al_h)}(u)R_{a,0}^{(\al_h,\al_i)}(u+\rmi u_0)\label{mondrom1}\\
\ov T(u)&=&\ov R_{a,L}^{(\al_h,\al_h)}(-u)\ov R_{a,L-1}^{(\al_h,\al_h)}(-u)\ldots \ov
R_{a,1}^{(\al_h,\al_h)}(-u)\ov R_{a,0}^{(\al_h,\al_i)}(-u+\rmi u_0)\nn\; ,
\ee
where indices $a$ ($\nu=0,\ldots,L$) refer to the auxiliary (quantum) spaces.
Note that the quantum space of the last factor (impurity) carries a representation different from that in the host, $\al_i\neq \al_h$. For the model \refeq{mod}, one chooses $\al_i=0$, $\al_h\equiv \al$. A shift by $\rmi u_0$ of the spectral parameter is allowed on the impurity site. By taking the supertrace over the auxiliary spaces of $T,\,\ov T$, one obtains the transfer matrices $\tau,\,\ov \tau$, giving rise to the Hamiltonian $H$, 
\be
\tau(u)&=& \str_a T(u) \; , \qquad \ov \tau(u)=\str_a \ov T(u)\label{bartaudef}\\
\ln \l[\tau\ov\tau\r](u)&=& \ln
\l[\tau\ov\tau\r](0)+u\underbrace{\l[\tau^{-1}(0)\tau'(u)+\ov\tau^{-1}(0)\ov\tau'(u)\r]_{u=0}}_{\displaystyle{=:{\rm
    const.
    } H}}+\Or(u^2)\label{defham}\; .
\ee  
One is free to choose the multiplicative constant in the definition of $H$ by scaling $u$. Here, we take $D(\al_h+1)$ as a common prefactor, where $D$ is a bandwidth parameter. External fields $\mu,h$ are included by twisted boundary conditions of the quantum spaces, for details on the calculation of $H$, see \cite{bo04}. 

Consider the quantum transfer matrix $\tau_{i,h}^{(Q)}$ pertaining to the impurity/host sites, \cite{bo04}, with Trotter number $N$. The free energy contributions of the impurity and host are given by the largest eigenvalue $\Lambda_{i,h}^{\max}$ of the QTM in the limit of infinite Trotter number, 
\be
f_i=-\lim_{N\to\infty}\frac{1}{\beta} \ln \Lambda_{i}^{\max}(u_0)\;,\;\;f_h=-\lim_{N\to\infty}\frac{1}{\beta} \ln \Lambda_{h}^{\max}(0)\label{ffromlam}\; .
\ee 
All eigenvalues of $\tau_{i,h}^{(Q)}$ can be found by ABA-techniques similar to \cite{frahm,goe02,grun99}, so that we do not repeat the calculation here. The result is
\be
\L_i^{(Q)}(v)&=& \phi_1(v)\,\frac{q_1\l(v+\frac\rmi2+\rmi\al_h\r)}{q_1\l(v+\frac\rmi2\r)}\,\te^{2\beta\mu}\nn\\
& &+\phi_2\l[\frac{q_1\l(v+\rmi\al_h+\frac\rmi2\r)}{q_1\l(v+\frac\rmi2\r)}\,\frac{q_2(v+\rmi)}{q_2(v)}\,\te^{\beta h/2}+\frac{q_1\l(v+\rmi\al_h+\frac\rmi2\r)}{q_1\l(v-\frac\rmi2\r)}\,\frac{q_2(v-\rmi)}{q_2(v)}\te^{-\beta h/2}\r]\,\te^{\beta\mu}\nn\\
& & +\phi_3(v)\,\frac{q_1\l(v+\rmi\al_h+\frac\rmi2\r)}{q_1\l(v- \frac\rmi2\r)}\label{liq}\; .
\ee
The vacuum expectation values are
\be
\phi_1(v)&=& \frac{\phi_+\l(v-\rmi\frac{\al_h}{2}+\rmi\frac{\al_i}{2}\r)\,\phi_+\l(v+\rmi-\rmi\frac{\al_h}{2}+\rmi\frac{\al_i}{2}\r)\,\phi_-\l(v-\rmi\frac{\al_h}{2}-\rmi\frac{\al_i}{2}\r)}
{\phi_+\l(v-\rmi\frac{\al_h}{2}-\rmi\frac{\al_i}{2}\r)\,\phi_+\l(v+\rmi+\rmi\frac{\al_h}{2}+\rmi\frac{\al_i}{2}\r)\,\phi_-\l(v+\rmi\frac{\al_h}{2}+\rmi\frac{\al_i}{2}\r)}\nn\\
\phi_2(v)&=&\frac{\phi_+\l(v-\rmi\frac{\al_h}{2}+\rmi\frac{\al_i}{2}\r)\,\phi_-\l(v-\rmi\frac{\al_h}{2}+\rmi\frac{\al_i}{2}\r)}{\phi_+\l(v+\rmi+\rmi\frac{\al_h}{2}+\rmi\frac{\al_i}{2}\r)\,\phi_-\l(v+\rmi\frac{\al_h}{2}+\rmi\frac{\al_i}{2}\r)}\nn\\
\phi_3(v)&=&\l.\phi_1(-v)\r|_{\al_{i,h}\to-\al_{i,h}-1}\nn\; ,
\ee
where $\phi_\pm(v):=\l(v\pm \rmi u\r)^{N/2}$ and $u:=-D(\al_h+1)\beta/N$. The $q$-functions are polynomials with zeroes at the BA-numbers, $q_\nu(v):=\prod_{k=1}^{M_\nu}\l(v-v_k^{(\nu)}\r)$, where $\nu=1,2$. The two sets of BA-numbers $\l\{v_k^{(1)}\r\}$, $\l\{v_k^{(2)}\r\}$ are determined by the analyticity of $\L_i^{(Q)}(v)$, 
\be
\frac{q_2\l(v_k^{(2)}-\rmi\r)\,q_1\l(v_k^{(2)}+\frac\rmi2\r)}{q_2\l(v_k^{(2)}+\rmi\r)\,q_1\l(v_k^{(2)}-\frac\rmi2\r)}\,\te^{-\beta h}&=&-1\nn\\
\frac{q_2\l(v_k^{(1)}+\frac\rmi2\r)}{q_2\l(v_k^{(1)}-\frac\rmi2\r)}\,\te^{\beta(h/2-\mu)}&=& -\frac{\phi_1\l(v_k^{(1)}-\frac\rmi2\r)}{\phi_2\l(v_k^{(1)}-\frac\rmi2\r)}\nn\; .
\ee
The eigenvalues of $\L_h^{(Q)}$ of the host matrix are given by an expression similar to \refeq{liq}, with $\al_i\equiv \al_h$. 

The free energy follows from the largest eigenvalue, determined by $M_\nu=N/2$ many BA-roots in both sets $\nu=1,2$. For this case, $\ln \L_h^{\max}(v)$ has been calculated in \cite{jue97,sak01} from appropriately chosen auxiliary functions, obeying a closed set of non-linear integral equations (NLIE). Within this approach, the Trotter limit $N\to\infty$ can be carried out analytically. It turns out that $\ln \L_i^{\max}(v)$ can be calculated analogously to $\ln \L_h^{\max}(v)$ with the result:
\be
\lim_{N\to\infty}\ln \L_i^{\max}(v)&=&\eta(v)+[\zeta*\ln \B](v)+[\ov\zeta*\ln \BB](v)+[(\zeta+\ov\zeta)*\ln \C](v)\label{impev}\;,
\ee  
where the auxiliary functions $\B=1+\b$, $\BB=1+\bb$ and $\C=1+\c$ obey the NLIE
\be
\ln \b(v)&=&\phi_\b(v+\rmi\delta) -[k_\b*\ln\BB](v+2\rmi\delta)-[k_\b*\ln\C](v+\rmi\delta)+\beta(\mu+h/2)\label{beqv2}\\
\ln \bb(v)&=&\phi_\bb(v-\rmi\delta) -[k_\bb*\ln\B](v-2\rmi\delta)-[k_\bb*\ln\C](v-\rmi\delta)+\beta(\mu-h/2)\label{bbeqv2}\\
\ln\c(v)&=&\phi_\c(v)
-[k_\b*\ln\BB](v+\rmi\delta)-[k_\bb*\ln\B](v-\rmi\delta)-[k_\c*\ln\C](v)+2\beta\mu\label{ceqv2}\; .
\ee
Here, the driving terms
\be
\phi_\b(v)&=&-\,\frac{\beta D(\alpha_h+1)^2}{(v+\rmi\al_h/2)(v-\rmi\al_h/2-\rmi)}\;,\;\phi_\bb=\phi_\b^*\label{phib}\\
\phi_\c&=&\phi_\b+\phi_\bb\label{phic}
\ee
and convolutions $[f*g](v):=\i f(v-w)g(w)\d w$ with integration kernels
\be
k_\b(v)&=&\frac{1}{2\pi v(v-\rmi)}\, ,\; k_\bb=k_\b^*\,,\;k_\c=
k_\b+k_\bb\nn\\
2 \pi \zeta(v)&=& -\l.\frac{\phi_\b(-v)}{D\beta(\al_h+1)}\r|_{\al_h\to\al_i}\, ,\; \eta(v)= 2\beta
D\frac{(\al_h+1)(\al_i/2+\al_h/2+1)}{v^2+(\al_i/2+\al_h/2+1)^2}\nn\; .
\ee
have been defined. The external fields enter through the asymptotic values of
the auxiliary functions,
\be
\b(\pm \infty)&=&\frac{\te^{\beta(\mu+h/2)}}{1+\te^{\beta(\mu-h/2)}}\;,\bb(\pm
    \infty)=\frac{\te^{\beta(\mu-h/2)}}{1+\te^{\beta(\mu+h/2)}}\label{asb}\\
\c(\pm \infty)&=& \frac{\te^{2\beta \mu}}{1+\te^{\beta (\mu+h/2)}+\te^{\beta
    (\mu-h/2)}}\label{asc}\; .
\ee
From \refeq{ffromlam} and \refeq{impev}, the free energy contribution of the
impurity results into
\be
f_i&=&\l\{\begin{array}{cc}
-T\eta\l(u_0\r)-T[\zeta*\ln \B]\l(u_0\r)-T[\ov\zeta*\ln \BB]\l(u_0\r)-T[(\zeta+\ov\zeta)*\ln
  \C]\l(u_0\r),&\al_i\neq 0\\
T\ln \c\l(u_0\r)+4 D(\al_h+1)\frac{\al_h}{4 u_0^2+\al_h^2}-2\mu,& \al_i=0.
\end{array}\r.\label{fimp}
\ee
The free energy of the host per lattice site is also given by
eq. \refeq{fimp}, with $\al_i\to\al_h$,
\be
f_h&=& \l.f_i\r|_{\al_i\to\al_h; u_0=0}\label{fhdef}.
\ee

Note that, apart from the $\eta$-term in \refeq{impev}, the parameter $\al_h$
enters explicitly only in the NLIE \refeq{beqv2}-\refeq{ceqv2}, whereas
$\al_i$ is found only in the free energy equation \refeq{fimp}. This had to be
expected: The quantum spaces of $\tau^{(Q)}_i$ are modules of the
$\al_h$-representation of gl(2$|$1). The auxiliary space, over which the trace has to be taken to get $\tau_i^{(Q)}$, carries the $\al_i$-representation. Especially, the auxiliary functions depend only on $\al_h$, and thus are identical for the host and impurity eigenvalues. However, the functional dependence of the eigenvalues on the auxiliary functions is different for the host and the impurity. 

The NLIE \refeq{bv3}-\refeq{cv3} are obtained from \refeq{beqv2}-\refeq{ceqv2} by setting $\delta=1/2$, $\al_h\equiv \al$, $\al_i=0$ and by making the substitution
\be
\l[k_{\b,\c}*\ln\C\r](v)&=&\l[k_{\b,\c}*\ln \c\r](v)+\l[k_{\b,\c} *\ln \frac{\C}{\c}\r](v)\nn.
\ee
Manipulations in Fourier space yield the result \refeq{bv3}-\refeq{cv3}. For $\al_i\neq 0$, eq. \refeq{fimp} yields the free energy of a four-site impurity, where the charge fluctuation terms include transitions from and to the doubly occupied state.

%%%%%%%%%%%%%%%%%%%
%%Appendix B
%%%%%%%%%%%%%%%%%%
\section{Appendix B}
\label{appb}
We calculate the contribution $\Or(T^2)$ to $f_h$, eq. \refeq{fhdef}. To this end the NLIE are linearized for $\beta D,\beta\mu,\beta \gg 1$, following a similar procedure applied to the homogeneous $tJ$-model in \cite{jue97tj}. Since $u_0=0$, the form \refeq{beqv2}-\refeq{ceqv2} is best suited for our analysis. First observe that for $h$ sufficiently large, $\bb=\Or\l(\te^{-\beta h}\r)$ and therefore $\BB=1$ up to exponentially small corrections. Secondly, both $\ln \b$, $\ln \c$ have two zeroes, $\ln \b(\pm \L_\b)=0$ and $\ln \c(\pm \L_\c)=0$. Due to the driving terms in \refeq{beqv2}-\refeq{ceqv2} of order $\Or(\beta)$, the slopes of the auxiliary functions in their zeroes behave as $|\ln_{\b,\c}'(\pm\L_{\b,\c})|\sim \beta$, so that there is a sharp crossover for $\beta\to\infty$ in the region around $\pm\L_{\b,\c}$. Thus we define the scaling functions $\ve_{\b,\c}$ by $\beta \ve_\b:=\ln\b$, $\beta \ve_c:=\ln \c$. Then
\be
\ln \B(v)=\beta \ve_\b(v)\theta\l(v^2-\L_\b^2\r)+\frac{\pi^2}{12\beta |\ve_\b(\L_\b)|}
\l(\delta(v-\L_\b)+\delta(v+\L_\b)\r)\label{bscale}\;,
\ee
and similarly for $\ln \C$, \cite{jue97tj, sak01}. Here, $\theta(v)$ denotes
the Heaviside function. Higher order terms can be
neglected if the slope is sufficiently steep. This approximation is inserted
into \refeq{beqv2}-\refeq{ceqv2} with $\delta=1/2$, such that the functions
are real-valued on the real axis. With $\k_{\b,\c}(v):=k_{\b,\c}(v+\rmi/2)$, this gives
\begin{subequations}
\be
\ve_\b(v)&=&\ve_{0,\b}(v)-\int_{|w|>\L_\c} \k_\b(v-w)\ve_\c(w)\,\d w\label{eb}\\
\ve_\c(v)&=&\ve_{0,\c}(v)-\int_{|w|>\L_\b} \k_\b(v-w)\ve_\b(w)\,\d w-\int_{|w|>\L_\c} \k_{\c}(v-w)\ve_\c(w)\,\d w\label{ec}\;,
\ee
\end{subequations}
where we have defined the dressed energy terms
\be
\ve_{0,\b}(v)&:=& \frac{\phi_\b(v+\rmi/2)}{\beta}+\mu+\frac h2-\frac{\pi^2}{12\beta^2|\ve_\c'(\L_\c)|}\l[\k_\b(v-\L_\c)+\k_\b(v+\L_c)\r]\nn\\
\ve_{0,\c}(v)&:=& \frac{\phi_\c(v)}{\beta}+2\mu-\frac{\pi^2}{12\beta^2|\ve_\b'(\L_\b)|}\l[\k_\b(v-\L_\b)+\k_\b(v+\L_b)\r]\nn\\
& &-\frac{\pi^2}{12\beta^2|\ve_\c'(\L_\c)|}\l[\k_\c(v-\L_\c)+\k_\c(v+\L_c)\r]\nn\; .
\ee
Eqs. \refeq{eb}, \refeq{ec} are written compactly as
\be
\l(\begin{array}{cc} 
1    & \k_\b\\
\k_\b &1+\k_\c
\end{array} \r)*
\l(\begin{array}{c}
\ve_\b\\
\ve_\c
\end{array}\r)=
\l(\begin{array}{c}
\ve_{0,\b}\\
\ve_{0,\c}
\end{array}\r)\label{dress1}\;,
\ee
where the convolutions are to be taken in the appropriate limits. Define two further functions $\xi_{\b,\c}$ by
\be
\l(\begin{array}{cc} 
1    & \k_\b\\
\k_\b &1+\k_\c
\end{array} \r)*
\l(\begin{array}{c}
\xi_\b\\
\xi_\c
\end{array}\r)=
\l(\begin{array}{c}
f_{\b}\\
f_{\c}
\end{array}\r)\nn\;,
\ee
with the same integration limits as in eq. \refeq{dress1} and 
\be
f_\b&:=&\l\{\begin{array}{ll}
          \l.\zeta\r|_{\al_i\to\al_h},&\alpha_h\neq 0\\
	  \k_\b, &\al_h =0
	  \end{array}\r.\nn\\
f_\c&:=&\l\{\begin{array}{ll}
          \l.\zeta+\ov\zeta\r|_{\al_i\to\al_h},&\alpha_h\neq 0\\
	  \k_\c, &\al_h =0.
	  \end{array}\r.
\ee
It follows that
\be
\lefteqn{\int_{|v|>\L_\b}\ve_\b(v)f_\b(v)\,\d v+\int_{|v|>\L_\c}\ve_c(v)f_\c(v)\,\d v}\nn\\
&=&\int_{|v|>\L_\b}\xi_\b(v)\ve_{0,\b}(v)(v)\,\d v+\int_{|v|>\L_\c}\xi_\c(v)\ve_{0,c}(v)\,\d v\label{dresbar} \; .
\ee
Writing the impurity contribution to the free energy for $u_0=0$, the second
case in \refeq{fimp}, in terms of the scaling functions \refeq{bscale}, and
using \refeq{dresbar}, one obtains
\be
f_h&=& -T \eta(0)+\l(\mu+\frac h2\r)\int_{|v|>\L_\b}\xi_\b(v)\,\d v+2\mu \int_{|v|>\L_\c}\xi_\c(v)\,\d v\nn\\
& &-T\int_{|v|>\L_\b} \phi_\b(v+\rmi/2)\xi_\b(v)\,\d v-T \int_{v>\L_\c} \phi_\c(v) \xi_\c(v)\,\d v\nn\\
& & -T^2\l\{\frac{\pi^2\xi_\c(\L_\c)}{6|\ve'_\c(\L_\c)|}+\frac{\pi^2\xi_\b(\L_\b)}{6 |\ve'_\b(\L_\b)|}\r\}\label{ttlim}\; .
\ee
This result reflects that at low
temperatures, the contributions of charge and spin excitations can be separated, but occur on the same temperature scale. The charge and spin velocities are given by
\be
\frac{1}{v_c}=\frac{\xi_\c(\L_\c)}{|\ve'_\c(\L_\c)|}\,;\;\frac{1}{v_s}=\frac{\xi_\b(\L_\b)}{|\ve'_\b(\L_\b)|}.\nn
\ee
For the explicit calculation of $v_{c,s}$ in certain limiting cases in the
range of $\L_{\b,c}$, one may
use Wiener-Hopf techniques; this has been done for $\al_i=\al_h=0$ \cite{sch87} and
$\al_h=0$, $\al_i\neq 0$ \cite{frahm}. The calculations of this appendix also apply to the $D\beta \gg 1$-behavior of $f_i$, if $u_0=0$. This demonstrates that the shift $u_0\sim \ln D$ induces a new energy scale which {\em energetically} separates charge from spin excitations.

\bibliographystyle{amsplain}

\providecommand{\bysame}{\leavevmode\hbox to3em{\hrulefill}\thinspace}
\providecommand{\MR}{\relax\ifhmode\unskip\space\fi MR }
% \MRhref is called by the amsart/book/proc definition of \MR.
\providecommand{\MRhref}[2]{%
  \href{http://www.ams.org/mathscinet-getitem?mr=#1}{#2}
}
\providecommand{\href}[2]{#2}

%\bibliography{../LIT/lit}
\end{document}